# Progressive Learning for Stabilizing Label Selection in Speech Separation with Mapping-based Method

*Chenyang Gao[1], Yue Gu[1], Ivan Marsic[1]*

[1] Department of Electrical and Computer Engineering, Rutgers University, Piscataway, NJ, USA

`{cg694,yg202,marsic}@rutgers.edu`

## Abstract

Speech separation has been studied in time domain because of lower latency and higher performance compared to time-frequency domain. The masking-based method has been mostly used in time domain, and the other common method (mapping-based) has been inadequately studied. We investigate the use of the mapping-based method in the time domain and show that it can perform better on a large training set than the masking-based method. We also investigate the frequent label-switching problem in permutation invariant training (PIT), which results in suboptimal training because the labels selected by PIT differ across training epochs. Our experiment results showed that PIT works well in a shallow separation model, and the label switching occurs for a deeper model. We inferred that layer decoupling may be the reason for the frequent label switching. Therefore, we propose a training strategy based on progressive learning. This approach significantly reduced inconsistent label assignment without added computational complexity or training corpus. By combining this training strategy with the mapping-based method, we significantly improved the separation performance compared to the baseline.

## 1. Introduction

Audio separation techniques are needed in practical applications of automatic speech recognition (ASR) [2]. In real-world communication, the speech usually is mixed with other signals, such as ambient sounds, reverberation, or other people's speech, and these signals may decrease the quality of the speech and dramatically reduce the performance of ASR. Speech separation aims to approximate the clean source for each speaker, which is needed for a complete ASR system.

Recent studies showed that time-domain methods have a higher performance and lower latency compared to time-frequency (T-F) domain methods [3, 4]. The time-domain speech separation with adaptive front-end was first proposed in [3], where it used 1-D convolution and 1-D transposed convolution to replace the STFT and iSTFT in T-F domain methods. A variety of models based on this convolutional encoder-separator-convolutional decoder structure have been proposed [4-8]. These approaches are all based on the masking-based method in the time domain, which aims to find a mask (weighted function) that could linearly apply to the representation of mixture waves to reconstruct the original clean sources. The other method [9, 10], the mapping-based method aims to approximate the clean sources directly from the mixture. However, their results showed the mapping-based method performed comparably or slightly worse on small datasets in the clean scenario. We extended the mapping-based study to a more challenging dataset [1], which allowed us to study the performance of masking- and mapping-based methods on different data sizes. Our ablation study on different-size training sets showed that mapping-based outperformed masking-based method on a large training corpus.

Permutation invariant training (PIT) [11] is an established approach to deal with label permutations in supervised speech separation tasks. PIT traverses all pairwise label-prediction permutations and selects the optimal one during the training. Although the PIT achieved great success in supervised speech separation, its underlying drawbacks still hinder a better separation quality [12-14], which is known as the inconsistent label selection problem. This inconsistent label assignment is caused by the small difference of the pairwise loss in the initial epochs [12-14]. We extended the study of this problem and found that the problem is more likely to be caused by the insufficient training of the model because the PIT worked well on a shallow model (with only one layer in the separator). Based on this observation, we determined why the multi-scale loss proposed in [15] improves the model training, which was left unexplained in [15]. However, the label selection was still inconsistent when using multi-scale loss as observed in the experiments. We then proposed a training strategy based on progressive learning that broadcasts the label assignments from initial to final, which effectively reduced the label switching among training epochs without additional computational complexity or extra data. By further combining the mapping-based model, we significantly improved the SI-SDR compared to the masking-based method trained only with PIT.

Our paper contributes:

1. We demonstrate that the mapping-based method obtained a better performance compared to the masking-based method using a large dataset.
2. We empirically analyzed the label selection problem and found that it was more likely under insufficient training of the model.
3. We proposed an efficient training strategy based on progressive learning that helped broadcast the label assignments across the model. Our proposed method significantly reduced the label permutation and led to better performance.

## 2. Effect of Data Size on Masking- and Mapping-based Methods

Although the mapping-based method has been studied [9, 10], these studies have been limited to training with limited-size datasets. The results showed a marginal improvement or even slightly worse performance in the clean scenario. We extended the training of the mapping-based method to a much larger training corpus to evaluate its generalization.

## 2.1. Masking-based versus Mapping-based approaches

The models based on TasNet [3] share a common structure: (1) an adaptive front-end that consists of a convolutional encoder and decoder, which replace the roles of STFT and iSTFT respectively, and (2) a separator that acts as its name suggests.

The problem of speech separation is to recover the clean source for each speaker from a given mixture as defined:

$$X = \sum_{1}^{c} s_i + n, \ s_i \in \mathbb{R}^t, \quad (1)$$

where $X$ is a mixture of the waveforms, $s_i$ is the clean source for speaker $i$, $c$ is the number of speakers, and $n$ is the background noise. The masking-based method aims to estimate a weighted function applied to the mixture so that:

$$[M_1; ...; M_c] = \text{Separator}(X)$$
$$\tilde{s}_i = M_i \odot X(t), \quad (2)$$

where $\tilde{s}_i$ is the reconstructed source, $M_i$ is the estimated mask for the source $i$ generated from the separation model, $\odot$ denotes the element-wise multiplication.

In the mapping-based approach, the separation model aims to approximate the target source directly from the mixture:

$$[\tilde{s}_1; ...; \tilde{s}_c] = \text{Separator}(X), \quad (3)$$

The mapping-based method omits the masking operation and performs better in the noisy scenario [10].

## 2.2. Separator

Variations based on Dual-Path RNN [6], including [5, 7, 8, 16], were proposed for the separation function. These approaches used dual-path processing [6], which segments long audio into short segments, then applies the intra and inter procedure sequentially. The dual-path processing allows the model to handle the long-range dependencies in long audios. We used an adaption of DPTNet [5] as a separation model. The basic block of DPTNet, the Improved Transformer [5], consists of two sublayers: a multi-head attention layer and an improved feedforward layer. We removed the ReLU [17] since the LSTM [18] has a lot of non-linearity. We used a modified DPTNet that we call DPTNet* as the baseline model in our experiments.

## 2.3. Dataset

Previous experiments with the mapping-based method used either the WSJ02MIX dataset [19] or a custom-made dataset [10]. Both datasets were relatively small. We extended these experiments to a more general setting, using the open-source LibriMix dataset [1] which was created using different subsets of the Librispeech dataset [20]. We used the Libri2Mix clean subset with a sample rate of 8k which consists of two-speaker mixtures. Train-100 (58 hours) and train-360 (212 hours) are used independently to measure the impact of the data scale on the performance of the methods.

## 2.4. Experiment setup

We implemented the model based on the Asteroid toolkit [21]. We used a window size of 16 and stride of 8 in convolutional encoder and decoder in all experiments. We used the default model configuration described in [4-6]. ReLU was used as an activation function for encoder and separation model output for the masking-based method to guarantee the non-negativity for both encoder output and mask. We omitted these activation functions in the mapping-based method because non-negativity is not needed. We used 200 training epochs with a batch size of 24 in all experiments. Adam [22] was used as an optimizer with an initial learning rate of 1e-3. The gradient is clipped with a maximum $L_2$ norm of 5. We reduced the learning rate if the validation loss could not improve in 5 consecutive epochs. The audios were segmented to 3 seconds during training and validation. We used SI-SDR (scale-invariant signal-to-distortion) [23] as a training objective in all the experiments. We compared the improvement of signal fidelity including SI-SDR improvement (SI-SDRi) and SDR improvement (SDRi).

Table 1: *Performance on Libri2Mix 100 and 360 subsets. SI-SDRi and SDRi (dB) are reported*

| Model | train-100 | train-360 |
|---|---|---|
| ConvTasNet [1] | 13.0/13.4 | 14.7/15.1 |
| DPRNN | 14.6/15.0 | 15.5/15.8 |
| DPTNet | 14.9/15.4 | 15.4/15.8 |
| DPTNet*-masking | **15.2/15.6** | 15.9/16.3 |
| DPTNet*-mapping | 14.9/15.0 | **16.3/16.7** |

## 2.5. Results of preliminary experiments

We first compared our modified DPTNet* with other baseline models with the masking-based method. All the dual-path-based models performed better than ConvTasNet (see Table 1). DPTNet performed better than DPRNN on a smaller training set (train-100) but did worse on a larger training set (train-360). The DPTNet*-masking consistently achieved the best results in both train-100 and train-360 compared to other models, showing that removing the ReLU activation in the improved transformer produced better performance. We observed that the mapping-based model performed worse than the masking-based one on a small dataset (which has a similar size as in [9, 10]). However, when we increased the training amount of data, the mapping-based method had around 0.4 dB performance improvement in SI-SDR compared to the masking-based method. This showed that mapping-based is more suitable for general scenarios since the training corpus could be arbitrarily enlarged by random mixing of clean input sources [24].

## 3. PIT and Label Switching Problem

Permutation invariant training (PIT) solved the problem of label selection in supervised learning of speech separation since the speaker order in the model output is unknown [11]. The PIT dynamically selects the optimal permutation by calculating the pairwise loss and uses this order to update the model. But, the learning is suboptimal because the label assignment may vary a lot across the training epochs [12-14]. That is, the optimal label permutation for the current epoch may be different from that of the previous epoch. They conjectured that the key reason for frequent label switching is that the difference of loss for different output-prediction pairs is small in the initial epochs, and PIT fails to keep the permutation selection [12-14]. This problem causes oscillations in the model updates, preventing a better performance. We extended the study of the label switching problem of PIT and determined that it is caused by insufficient training of the separation model, as described next.

### 3.1. Comparing shallow and deep models

To show that the label switching is caused by the unregularized training of the model, we used two modified DPTNets* trained with the original PIT. These models differed only in the number of layers in the separation model, where the shallow model had only one block ($B=1$) and the deep model had six blocks ($B=6$), as in previous studies. We stored the model in each epoch and analyzed how the label assignment for training data switched across the training epochs. Previous studies [12-14] compared the label assignments in the successive epochs, which obscured a comparison of label assignment switching in nonadjacent epochs. We compared the label assignment in each epoch to that for the best model obtained from the training—the one that performed best on the validation set (Figure 1). This approach allowed us to independently evaluate the label switching in each epoch compared to a standard baseline. Ideally, label switching compared to the best model should be 0 if label assignments were the same. The label assignments for the deep model switched frequently in the initial training epochs, then stabilized as we gradually reduced the learning rate (Figure 1). The label assignments from epochs 1-75 were almost opposite from those of the best model. These observations indicate that the deep model was trained in a different direction in early epochs vs. later epochs. On the other hand, the label assignment ratio for the shallow model was less than 10%, and the label assignments were almost identical to those of the best model. These findings showed that PIT works well for the shallow model, where the label switching occured much less than in the deep model. The fact that in initial epochs the label assignments do not change frequently, rejected the previous conjecture that the trivial loss difference between different pairs leads to the inconsistency in optimal label selection in the initial epochs [12-14].

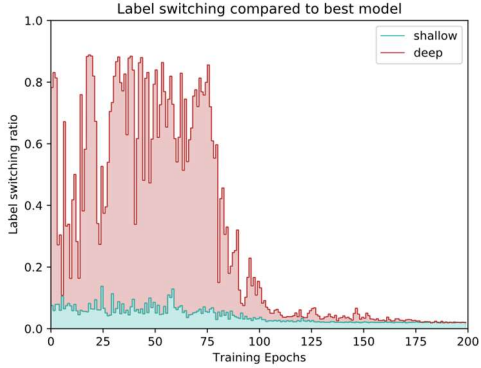

Figure 1: Label assignment switching accross the training epochs compared to the best model.

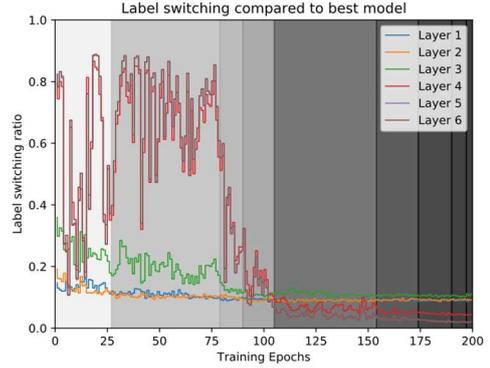

Figure 2: Label assignment switching in each layer of the deep model, compared to the label assignment in each layer of the best model.

### 3.2. Why deep model makes inconsistent label assignment

Although the PIT worked well with the shallow model, the reason for frequent label switching in the deep model remained unclear. We analyzed the label switching for different layers of the deep model (Figure 2). Different shades of grey in the background show where we halved our learning rate during training. We made several observations: (1) The label switching ratios of Layers 1-3 were correlated, and their label assignments did not change frequently compared to the final label assignments. The label switching ratios of Layers 4-6 were highly correlated (almost identical in Layers 5 and 6); their label assignments with the first three learning rates frequently switched compared to the final label assignments. (2) The label assignment switching should have converged after the learning rate decreased several times, and this is true for Layers 4-6 where the label switching diminished after epoch 125. However, the label switching in Layers 1-2 remained constant along the training epochs, unaffected by the learning rate reduction.

Layer 3 label switching dropped from the initial rate, then remained constant.

We call the phenomenon of dissimilar label switching across adjacent layers a "layer decoupling problem." Our experiments showed that Layers 4-6 were decoupled from Layers 1-3 (Figure 2). We conjectured that the separation performance was strongly affected by insufficient training of the initial layers (Layers 1-3), which led to the decoupling problem. This problem in turn causes the label switching problem of the deep model. On the other hand, a shallow model with a single layer did not suffer from the decoupling problem, and the separation model was adequately and correctly trained so that label switching was negligible in the shallow model.

## 4. Broadcasting Label Assignment Across Layers via Progressive Learning

A solution to insufficient training could stabilize the label assignments if it: 1) adequately trains the initial layers to avoid layer decoupling, and 2) broadcasts the label assignments across layers. Progressive learning has been applied in speech enhancement [25] and speech separation [15], which used multiple loss terms to force the model to progressively enhance or separate the speech across the layers of the model. However, the reason why this progressive learning-based loss helped to improve the separation performance was not identified in [15]. We first review their approach and offer our explanation for their observed improvement. We then present our progressive learning-based training method that outperforms their approach. Finally, we discuss and analyze the results of these two methods.

### 4.1. The multi-scale loss approach

Nachmani, et al. [15] proposed the multi-scale loss approach to speech separation based on progressive learning and showed a performance improvement. The idea of multi-scale loss is to require the model to separate the mixed waves in a progressive manner, by using the intermediate output from each layer to reconstruct the clean sources. It is defined as [15]:

$$Loss = \frac{1}{B}\sum_{i=1}^{B} \text{PIT}(\hat{S}_i, S) \qquad (4)$$

A pairwise loss is calculated after each layer, and the optimal label assignment is selected by PIT. The losses are added and used to update the model. The multi-scale loss regularizes the training of each layer, reducing the layer decoupling problem in the deep model. However, overregularization may prevent even better performance since it results in contradicting model training. Specifically, in a 6-layer model, the first layer is updated using six loss terms while the last layer is updated with one loss term. The initial layers may have suboptimal training because different loss terms may contradict each other in a single training sample. Our experiments showed that multi-scale loss outperforms the original PIT, but the layer decoupling problem remains due to the overregularization, leading to an unstable label assignment problem.

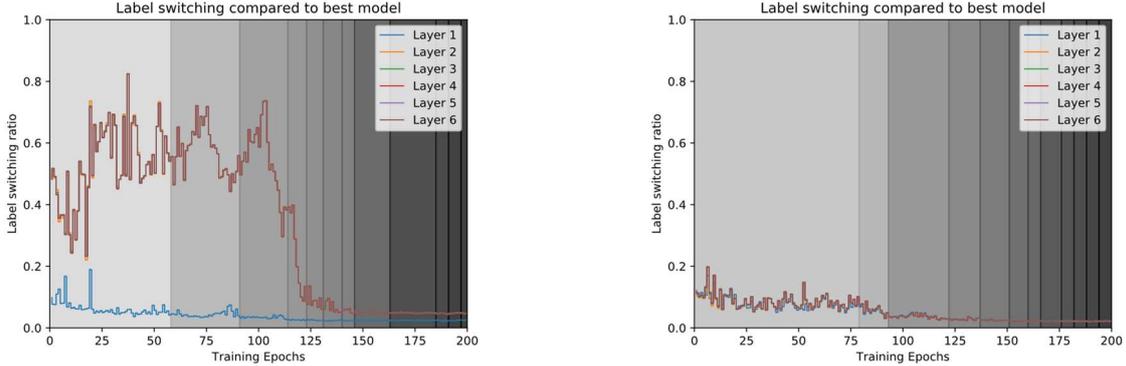

Figure 3: Label assignments switches in each layers compared to the best model (a) multi-scale loss (b) early-break.

### 4.2. Early-break progressive learning approach

We propose a variant of progressive learning called early-break progressive learning to address the problem of contradicting losses in the multi-scale loss approach [15]. Consider a TasNet with $B$ layers in the separation model. During the forward pass of training, we select a layer index between 1 to $B$ as an early-break layer. We use the output from the early-break layer and ignore the remaining layers. Then we use PIT to train the layers up to the early-break layer. This approach has two advantages in speech separation compared to the multi-scale loss approach. First, the early break allows the model to access the output from middle layers, which maintain the main property of progressive learning, but only the output from the early-break layer is used to calculate the pairwise loss, which effectively reduces the training time, unlike the multi-scale loss that requires $B$ times of pairwise comparison in PIT. Second, only one loss term is used for updating the model, which avoids overregularization caused by contradicting losses. The loss for our early-break progressive learning is defined as:

$$Loss = \text{PIT}(\hat{S}_i, S) \qquad (5)$$

where $i$ refers to a randomly-generated early break index during training. To ensure a stable training of the model, half of the early break indices were set to $B$ to train all the layers as in the original PIT. Our experiment results showed that early-break progressive learning addressed the layer decoupling problem, and broadcast the label assignments across the layers, which greatly reduced the unstable label assignment problem.

### 4.3. Results and analysis of the multi-scale loss approach

We used the multi-scale loss [15] for training the DPTNet* model. We trained our model on both train-100 and train-360 sets in LibriMix. The performance of the multi-scale loss improved for both methods compared to PIT, and the mapping-based method always performed better (Table 2).

To determine how the multi-scale loss improves over PIT and leads to better performance, we again analyzed the label assignments in mapping-based method across the training epochs for each layer. We found that the layer decoupling problem is greatly reduced (Figure 3(a)) because the multi-scale loss activates the training of some initial layers. Layers 2-6 now are coupled, thus improving the separation performance. However, the label assignments still frequently switch, as Layer 1 is still decoupled from other layers (Figure 3(a)). We conjecture that this is caused by the overregularized training (Section 4.1) and the label assignments cannot successfully broadcast across all layers due to the contradicting loss terms. This label switching still hindered a better performance in the multi-scale loss approach.

### 4.4. Results and analysis of the early-break approach

The same experiments were repeated to evaluate our early-break progressive learning approach. The separation performance of early-break progressive learning outperformed the models trained on PIT and the multi-scale loss (Table 2). Label switching analysis for early-break progressive learning is shown in Figure 4(b). Since only one loss term is used to update the model during training, this avoided the contradicting loss terms of the multi-scale loss. Now all layers in the model are tightly coupled, thus broadcasting the stable label assignments from initial layers to avoid suboptimal training.

## 5. Conclusion

We extended the study of masking- and mapping-based methods used in time-domain separation networks and the label switching problem in the permutation invariant training (PIT). Our ablation study showed that the mapping-based method worked better on a larger dataset, making it more suitable for real-world applications since arbitrarily large datasets can be simulated by dynamic mixing. Our analysis showed that the label switching problem is likely caused by insufficient training in the initial layers of separation models. We determined why the multi-scale loss approach improved the speech separation and identified its remaining limitations. We introduced early-break-based progressive learning that broadcasts stable label assignments in the initial layers and significantly reduces the label assignments switching that affected the original PIT.

Table 2: *The separation performance of masking and mapping-based methods; SI-SDRi and SDRi (dB) are reported.*

|  | train-100 | train-360 |
| --- | --- | --- |
| PIT-masking | 15.2/15.6 | 15.9/16.3 |
| PIT-mapping | 14.8/15.0 | 16.3/16.7 |
| Multi-scale-masking | 15.7/16.2 | 17.1/17.2 |
| Multi-scale-mapping | 15.9/16.3 | 17.3/17.7 |
| Early-break-masking | 15.9/16.3 | 17.2/17.6 |
| **Early-break-mapping** | **16.3/16.8** | **17.5/17.9** |